\documentclass[prb,superscriptaddress,showpacs,reprint,floatfix]{revtex4-1}

\usepackage{amsmath}
\usepackage{braket}
\usepackage{graphicx}
\usepackage{epstopdf}

\newcommand{\Hamil}{\hat{\mathcal{H}}}
\newcommand{\Sp}{\hat{S}}

\newcommand{\Ea}{\mathcal{E}_a}
\newcommand{\muB}{\mu_B}
\newcommand{\HT}{H_T}
\newcommand{\HA}{H_A}

\newcommand{\ie}{\emph{i.e.}}
\newcommand{\eg}{\emph{e.g.}}

\begin{document}

\title{Anisotropic properties of spin avalanches in crystals of nanomagnets}

\author{C. M. Dion}

\email{claude.dion@physics.umu.se}

\affiliation{Department of Physics, Ume{\aa} University, SE-901\,87 Ume{\aa},
  Sweden}

\author{O. Jukimenko}

\affiliation{Department of Physics, Ume{\aa} University, SE-901\,87 Ume{\aa},
  Sweden}

\author{M. Modestov}

\affiliation{Nordita, AlbaNova University Center, SE-106\,91
  Stockholm, Sweden}

\author{M. Marklund}
\email{mattias.marklund@physics.umu.se}

\altaffiliation[Also at: ]{Applied Physics, Chalmers University of
  Technology, SE-412\,96 G\"{o}teborg, Sweden} \affiliation{Department of
  Physics, Ume{\aa} University, SE-901\,87 Ume{\aa}, Sweden}

\author{V. Bychkov}
\email{vitaliy.bychkov@physics.umu.se}

\affiliation{Department of Physics, Ume{\aa} University, SE-901\,87 Ume{\aa},
  Sweden}

\begin{abstract}
  Anisotropy effects for spin avalanches in crystals of nanomagnets
  are studied theoretically with the external magnetic field applied
  at an arbitrary angle to the easy axis. Starting with the
  Hamiltonian for a single nanomagnet in the crystal, the two
  essential quantities characterizing spin avalanches are calculated:
  the activation energy and the Zeeman energy. The calculation is
  performed numerically for the wide range of angles and analytical
  formulas are derived within the limit of small angles. The
  anisotropic properties of a single nanomagnet lead to anisotropic
  behavior of the magnetic deflagration speed. Modifications of the
  magnetic deflagration speed are investigated for different angles
  between the external magnetic field and the easy axis of the
  crystals. Anisotropic properties of magnetic detonation are also
  studied, which concern, first of all, temperature behind the leading
  shock and the characteristic time of
  spin switching in the detonation.
\end{abstract}

\pacs{75.50.Xx, 75.45.+j, 75.60.Jk, 47.40.Rs}

\date{\today}

\maketitle

\section{Introduction}

Single-molecule magnets (nanomagnets) embedded in crystals are
compounds that exhibit unique physical properties with promising
applications to quantum computing and data
storage.\cite{mag:lis80,mag:caneschi91,mag:papaefthymiou92,mag:sessoli93b}
In particular, they can possess a large effective spin number
($S\sim10$)\footnote{For example, $\mathrm{Mn}_{12}$-acetate has a
  core of four $\mathrm{Mn}^{+4}$ ($S = 3/2$) surrounded by a ring of
  eight $\mathrm{Mn}^{+3}$ ($S = 2$), producing a ferromagnetic
  structure with a total spin number $S=10$, see, \eg,
  Ref.~\onlinecite{mag:thomas96}.} and show an anisotropy with
respect to the orientation of this spin, with the lowest energy
corresponding to an ``easy axis'' of the
crystal.\cite{mag:sessoli93,mag:villain94} That is, the potential
energy as a function of the orientation of the spin exhibits a
double-well structure, even in the absence of any external magnetic
field.  In the presence of an external magnetic field parallel to the
easy axis, the two wells are asymmetric and the spin aligns with the
field.  Upon a sudden reversal of the field, the internal crystal
anisotropy creates a barrier to the flip of the spin, and relaxation
may take place through \emph{spin
  tunneling}.\cite{mag:paulsen95,mag:friedman96,mag:thomas96,mag:garanin97,%
  mag:chudnovsky98,mag:villain00,mag:wernsdorfer01,mag:gatteschi03,%
  mag:delbarco05} It is also possible to trigger locally the
relaxation and, as it releases energy, observe the propagation of a
spin reversal front, corresponding to a magnetic
\emph{deflagration}\cite{mag:suzuki05,mag:hernandez-minguez05,mag:garanin07,%
  mag:modestov11a} or
\emph{detonation},\cite{mag:decelle09,mag:modestov11b} depending on
the speed and structure of the front. Magnetic deflagration and
detonation have much in common with the respective combustion
phenomena\cite{comb:law06,comb:bychkov00} (including the terminology);
there are even indications on the possibility of magnetic
deflagration-to-detonation transition similar to that studied
intensively within combustion
science.\cite{comb:bychkov05,comb:bychkov08,comb:dorofeev11}

Up to now, the research on magnetic deflagration and detonation has
mostly been restricted to unidimensional models, where the external
magnetic field is co-linear with the easy axis and the spin avalanche
front propagates along the same axis. Within such a restriction, one
obviously loses the possibility of an anisotropic spin interaction
with the magnetic field, together with an anisotropic propagation of
the avalanche fronts. Although the importance of and interest in the
anisotropic properties of spin avalanches was expressed from the very
beginning,\cite{sarachik-private} only a few papers addressed these
properties,\cite{mag:macia09,mag:velez12} which may be explained by
the experimental difficulties encountered in its study. In particular,
Ref.~\onlinecite{mag:macia09} investigated experimentally the
possibility of spin-avalanche initiation (``ignition'') for the
magnetic field inclined at an arbitrary angle to the easy axis. In
Ref.~\onlinecite{mag:velez12}, the authors compared the magnetic
deflagration speed for propagation along the easy axis (c) and the
hard axes (a or b) with the magnetic field collinear with the front
velocity vector. Thus, although the experimental data on the subject
is limited, the anisotropic properties of the magnetic deflagration
and detonation may be investigated using nanomagnet model
Hamiltonians.\cite{mag:delbarco05} To the best of our knowledge, no
theoretical investigation of these anisotropic properties has been
performed so far.  At the same time, the study of the anisotropic
properties gives a clue to the multidimensional dynamics of magnetic
deflagration and detonation. Multidimensional phenomena are known to
play the decisive role in traditional combustion
science;\cite{comb:law06,comb:bychkov00,comb:bychkov05,comb:bychkov08,comb:dorofeev11}
similar multidimensional pseudo-combustion effects have been also
obtained recently in advanced materials in the context of doping
fronts spreading in organic
semiconductors.\cite{OSC:bychkov11,OSC:modestov11,OSC:bychkov12}

In the present paper, we explore the effects of misalignment between
the external magnetic field and the easy axis.  We shall focus on the
development of a model for magnetic deflagration and detonation in a
crystal of single-molecule magnets in a generic magnetic field.  While
this model can be applied to any such system, specific calculations
will be based on Mn$_{12}$-acetate, which has an effective spin number
$S=10$.\cite{mag:lis80,mag:caneschi91,mag:sessoli93b} Starting with
the Hamiltonian for a single magnet embedded in the crystal, we
calculate the two essential quantities -- the \emph{activation energy}
and the \emph{Zeeman energy} -- characterizing the spin avalanche. We
investigate modifications of the magnetic deflagration speed produced
by misalignment of the magnetic field with the easy axis. We also
study the anisotropic properties of magnetic detonation, focusing on
the temperature behind the leading shock and for completed spin
reversal, and the characteristic time of spin switching. Unlike for
magnetic deflagration, the magnetic detonation speed is determined by
the sound speed and does not depend on the direction of the external
magnetic field.

The paper is organized as follows. We start by presenting, in the next
section, the quantum-mechanical calculation of the activation energy
and the Zeeman energy.  We then derive, in Sec.~\ref{sec:approx},
approximate analytical formulas for these values, based either on
quantum-mechanical perturbation theory or on a classical model for the
spin. In Sec.~\ref{sec:properties}, we consider the implications of
the quantum-mechanical results on magnetic deflagration and detonation
properties. Finally, we summarize our results in
Sec.~\ref{sec:summary}.

\section{Quantum-mechanical derivation of the activation and
  Zeeman energies}

\subsection{Hamiltonian for a single-molecule magnet}

A rather elaborate spin Hamiltonian for a molecular magnet, such as
Mn$_{12}$-acetate, can be written as\cite{mag:delbarco05}
\begin{align}
  \Hamil &= -D \Sp_z^2 - B \Sp_z^4 \nonumber \\
  & \quad - g \muB \left[ H_z \Sp_z + \HT \left(
      \cos \phi \Sp_x + \sin \phi \Sp_y \right)
  \right] \nonumber \\
  & \quad + E \left( \Sp_x^2 - \Sp_y^2 \right) + C \left( \Sp_+^4 +
    \Sp_-^4 \right) + \Hamil',
  \label{eq:hamilton}
\end{align}
with the spin raising and lowering operators $\Sp_\pm = \Sp_x \pm i \Sp_y$. The first two terms of Eq.~(\ref{eq:hamilton}) correspond to the uniaxial magnetic anisotropy,
while the third term is the interaction with a magnetic field
$\mathbf{H}$, oriented along the spherical angles $(\theta,\phi)$,
with the components
\begin{align}
H_x &\equiv H \sin \theta \cos \phi, \nonumber \\
H_y &\equiv H \sin \theta \sin \phi, \nonumber \\
H_z &\equiv H \cos \theta,
\end{align}
while
\begin{equation}
\HT \equiv \sqrt{H_x^2 + H_y^2}
\label{eq:HT}
\end{equation}
is the transverse magnetic field.  The 4th and 5th terms of
Eq.~(\ref{eq:hamilton}) are transverse anisotropy terms (inherent to
the molecule), and $\Hamil'$ contains additional terms due to the
inter-molecular dipole interaction and the hyperfine interaction with
the spin of the nuclei.  A set of values for the parameters in this
Hamiltonian for Mn$_{12}$-acetate can be found in
Tab.~\ref{tab:params}.
\begin{table}[b]
  \caption{\label{tab:params}Values of the different parameters of the
    spin Hamiltonian, Eq.~(\ref{eq:hamilton}), for Mn$_{12}$-acetate.}
  \begin{tabular}{clc} \hline
    Parameter & \multicolumn{1}{c}{Value} & Ref. \\ \hline
    $g$ & $1.93$ & [\onlinecite{mag:sessoli93b}] \\
    $D$ & $0.548$ K & [\onlinecite{mag:delbarco05}] \\
    $B$ & $1.17 \times 10^{-3}$ K & [\onlinecite{mag:delbarco05}] \\
    $E$ & $1.0 \times 10^{-2}$ K & [\onlinecite{mag:delbarco05}] \\
    $C$ & $2.2 \times 10^{-5}$ K & [\onlinecite{mag:delbarco05}] \\ \hline
  \end{tabular}
\end{table}

\begin{figure*}
  \centerline{\includegraphics{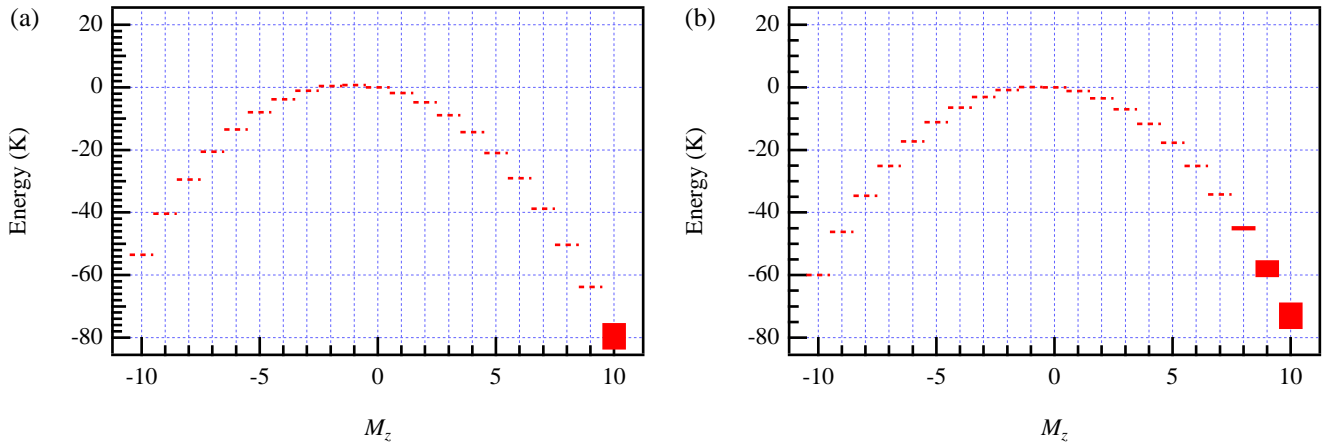}}
  \caption{\label{fig:levels}(Color online) Energies of the
    eigenstates of $\Sp_z$, labelled by the quantum number $M_z$, in
    an external field of 1\,T oriented along (a) $\theta = \phi = 0$;
    (b) $\theta = \pi/3$, $\phi = 0$.  The projection of the ground
    state on the different $M_z$ levels, $\left| \Braket{M_z | \psi_g}
    \right|^2$, is schematically represented by the thickness of the
    line (in a logarithmic-like scale), with dotted lines $\sim0$.}
\end{figure*}

\begin{figure}
  \includegraphics{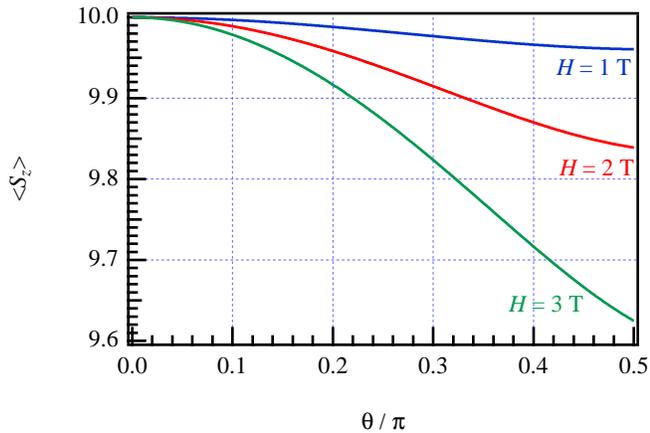}
  \caption{\label{fig:angle}(Color online) Expectation value of
    $\braket{\Sp_z}$ in the ground state $\Ket{\psi_g}$ of
    Hamiltonian~(\ref{eq:hamilton}) for Mn$_{12}$-acetate as a
    function of the polar angle $\theta$ between the magnetic field
    $\mathbf{H}$ and the easy anisotropy axis of the crystal, with
    $\phi=0$, for three different magnitudes of the field.}
\end{figure}

Even in the absence of a magnetic field, the presence of the
transverse anisotropy terms makes it such that the eigenstates of
$\Sp_z$ are not eigenstates of the full
Hamiltonian~(\ref{eq:hamilton}).  Nevertheless, due to the small
values of $E$ and $C$, it is still informative to discuss the problem
in terms of the magnetic quantum number $M_z$ associated with $\Sp_z$.
We plot in Fig.~\ref{fig:levels}(a) the energy of the $M_z$
eigenstates in a field of $H=1\, \mathrm{T}$ aligned along $z$ ($\theta
= \phi = 0$).
As $E$ and $C$ are small perturbations, the eigenvalues of $\Hamil$
are almost those of $\Sp_z$, and only the $M_z = 10$ level is
significantly present in the ground state.  Rotating the polar angle
to $\theta = \pi/3$ changes not only the energy of the $M_z$ levels,
Fig.~\ref{fig:levels}(b), but also increases the ``population'' of the
different $M_z$ in the ground state of the system, that is, the
projection of the ground state $\psi_g$ on the eigenstates of $\Sp_z$,
$\left| \Braket{M_z | \psi_g} \right|^2$.  While the ground state is
still located close to the maximum projection of the spin on the $z$
axis, \ie, the system is found in a single well of the double-well
structure, the energy of the ground state is higher than that of the
lowest $M_z$ level.  This can be observed by considering the
expectation value of $\braket{\Sp_z}$ in the ground state of
Hamiltonian~(\ref{eq:hamilton}) for different orientations of the
magnetic field, Fig.~\ref{fig:angle}.

The combination of the change of the level structure and the
projection of the initial and ground states on many levels will affect
the values of the activation and Zeeman energies, as described in
Sec.~\ref{sec:model}.  In all cases, we need the anisotropy to play
the dominant role, so that the double-well structure of the spin
energy is present.  Defining the anisotropy field
as\cite{mag:garanin97}
\begin{equation}
  \HA \equiv  (2S-1) \frac{D + B \left[ 2S \left( S-1 \right) + 1
    \right]}{g \muB}
  \label{eq:anisotropy}
\end{equation}
we must have at all times $H_z < \HA$ and $\HT \ll \HA$, with $\HA
\approx 11.1\, \mathrm{T}$ for Mn$_{12}$-acetate.  Also, while the
Hamiltonian~(\ref{eq:hamilton}) is different along $x$ and $y$, this
leads only to minimal modifications in the energy as $\phi$ is varied,
and we will thus concentrate on the behavior in the $xz$-plane, \ie,
for $\phi=0$.

\subsection{Determining the activation and Zeeman energies}
\label{sec:model}

The physical situation we consider here is the following.  Initially,
a crystal of molecular magnets is immersed in an external magnetic
field $\mathbf{H}_-$, which is then very rapidly inverted to a new
field $\mathbf{H} = -\mathbf{H}_-$.  Because of the magnetic
anisotropy, the system is then in a metastable state, and an energy
barrier must be overcome for it to relax to the new ground state.  The
relaxation of a given molecular magnet can then happen through spin
tunneling, where less energy than the barrier height is required, or
by thermal excitation above the barrier.  The molecular magnet thereby
releases the thermal energy equivalent to the difference in energy
between the initial metastable state and the actual ground state.
This thermal energy can then contribute to the relaxation of
neighbouring molecular magnets, hence the possibility of deflagration
and detonation inside the crystal.

In order to serve for the study of deflagration and detonation, our
model must therefore produce two main values, the \emph{activation
  energy} $\Ea$, \ie, the difference between the maximum energy of the
molecular magnet in the field $\mathbf{H}$ and the energy of the
initial metastable state, and the \emph{Zeeman energy} $Q$, corresponding
to the difference between the metastable state and the ground state in
the field $\mathbf{H}$.  Therefore, we first solve
\begin{equation}
   \Hamil_- \ket{\psi_i} = \mathcal{E}_{-,0} \ket{\psi_i},
\end{equation}
with $\Hamil_-$ the Hamiltonian using the field $\mathbf{H}_-$ (\ie,
the field before inversion), for $\mathcal{E}_{-,0}$ the lowest
eigenvalue of $\Hamil_-$, and then calculate the energy of that state
in the field $\mathbf{H}$,
\begin{equation}
\mathcal{E}_i = \braket{\psi_i | \Hamil | \psi_i}.
\end{equation}

To get the barrier height, we consider the spin-phonon coupling as a
sum over products of all the spin operators $\Sp_x$, $\Sp_y$, and
$\Sp_z$, (see, \emph{e.g.}, Ref.~\onlinecite{mag:koloskova63}), such
that the system overcomes the barrier by stepping through intermediate
states up to the state of highest energy $\mathcal{E}_\mathrm{max}$ in
the field $\mathbf{H}$,\cite{mag:gatteschi03} \ie,
\begin{equation}
  \Hamil \ket{\psi_\mathrm{max}} = \mathcal{E}_\mathrm{max}
  \ket{\psi_\mathrm{max}},
\end{equation}
such that
\begin{equation}
  \Ea = \mathcal{E}_\mathrm{max} - \mathcal{E}_i.
\end{equation}
Note that this model takes into account the effect of tunneling on the
position of the energy levels, but not the dynamical effects of
tunneling.  In other words, we consider that the crossing of the
barrier due to thermal excitation will be much faster than the
tunneling across it (opposite to what is studied in
Ref.~\onlinecite{mag:macia09}).

\begin{figure*}
  \centerline{\includegraphics{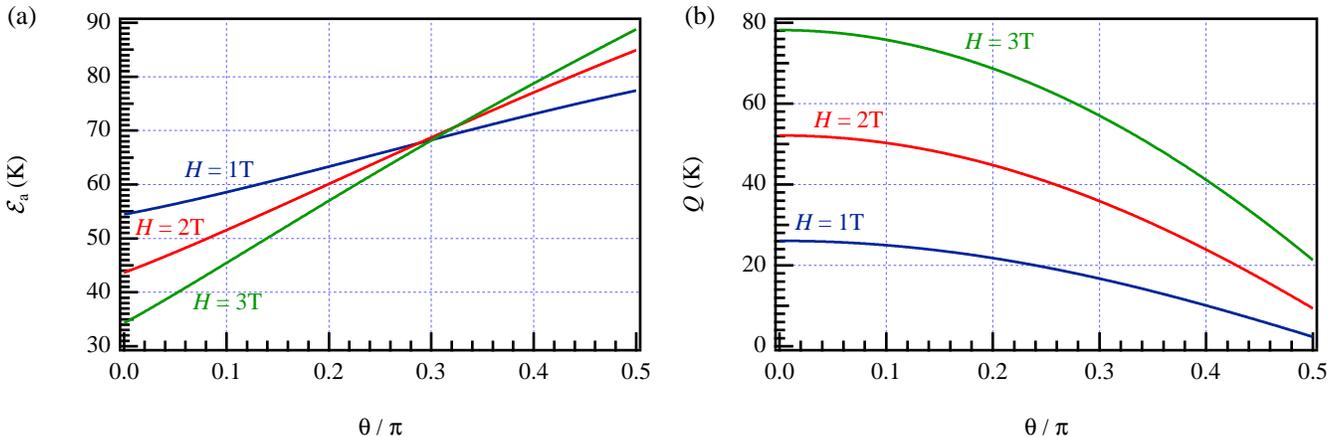}}
  \caption{\label{fig:Ea_Q}(Color online) Quantum-mechanical
    calculation of (a) the activation energy and (b) the Zeeman energy
    of Mn$_{12}$-acetate as a function of the polar angle $\theta$
    between the magnetic field $\mathbf{H}$ and the easy anisotropy
    axis of the crystal, with $\phi=0$, for three different magnitudes
    of the field.}
\end{figure*}

\begin{figure*}
  \centerline{
    \includegraphics{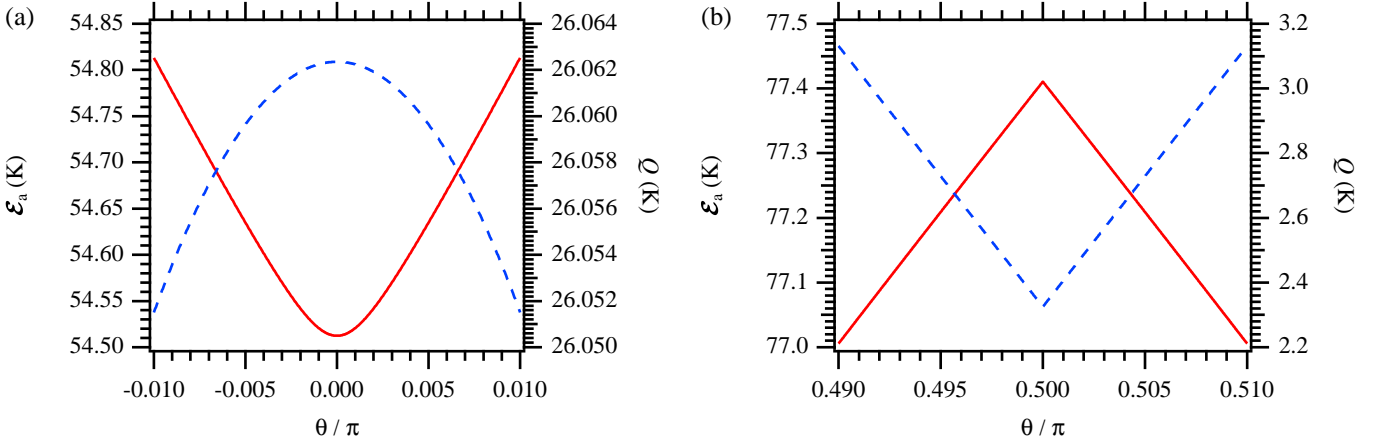}
}
  \caption{\label{fig:symmetry}(Color online) Quantum-mechanical
    calculation of the activation energy (solid line) and the Zeeman
    energy (dashed line) of Mn$_{12}$-acetate around the symmetry
    angles (a) $\theta=0$ and (b) $\theta = \pi/2$, for $\phi=0$ and
    $H= 1\, \mathrm{T}$.}
\end{figure*}

The Zeeman energy is itself found from the state of lowest energy
$\mathcal{E}_\mathrm{min}$ in the field $\mathbf{H}$,
\begin{equation}
  \Hamil \ket{\psi_\mathrm{min}} = \mathcal{E}_\mathrm{min}
  \ket{\psi_\mathrm{min}},
\end{equation}
as
\begin{equation}
  Q = \mathcal{E}_i - \mathcal{E}_\mathrm{min}.
  \label{eq:Q}
\end{equation}
Both $\Ea$ and $Q$ are easily calculated numerically, and some results
for a magnetic field in the $xz$-plane are presented in
Fig.~\ref{fig:Ea_Q}.
From the structure of the Hamiltonian~(\ref{eq:hamilton}), while it is
clear that these values are mirrored about $\theta=0$ and
$\theta=\pi/2$, there is a difference in behaviour of the curves
around these two angles.  While the Hamiltonian is symmetric about
both $\theta=0$ and $\pi/2$, the presence of $\HT$, Eq.~(\ref{eq:HT}),
makes the first derivative of the energy discontinuous at
$\theta=\pi/2$, and this is reflected in both $\Ea$ and $Q$, as can be
seen in Fig.~\ref{fig:symmetry}.

\subsection{Range of validity of the model}

\begin{figure}
  \centerline{\includegraphics{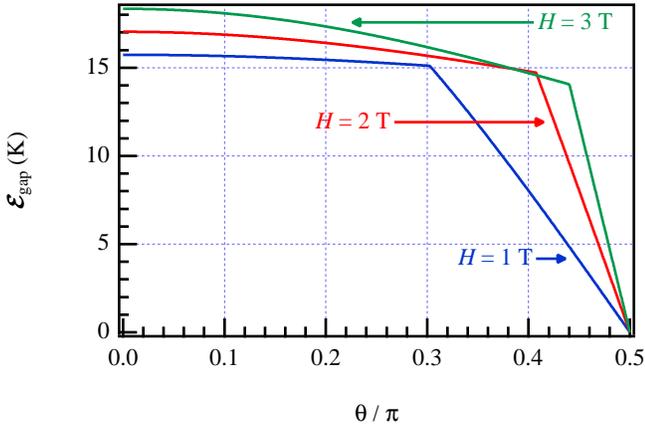}}
  \caption{\label{fig:Egap}(Color online) Energy gap between the
    lowest and next-to-lowest eigenstate of the
    Hamiltonian~(\ref{eq:hamilton}) for Mn$_{12}$-acetate as a
    function of the polar angle $\theta$ between the magnetic field
    $\mathbf{H}_-$ and the easy anisotropy axis of the crystal, with
    $\phi=0$, for three different magnitudes of the field.}
\end{figure}

An underlying assumption of this model is that, initially, a single
quantum level of the molecular magnet is populated.  This is of course
dependent on the initial temperature of the system, so it is useful to
also look at the difference in energy between the lowest state in
field $\mathbf{H}_-$ and the next-to-lowest.  We denote this quantity by
$\mathcal{E}_\mathrm{gap}$ and a plot of its value can be found in Fig.~\ref{fig:Egap}.
The curves clearly show a change of behavior for a certain value of
the angle $\theta$, which can be easily understood as follows.  If,
for the sake of the explanation, we neglect the fact that more than
one $M_z$ level is populated and only think in terms of the energies
of the $M_z$ states, for small angles the difference in energy
corresponds to that between $M_z=-10$ and $M_z=-9$ (for
$\mathbf{H}_-$, the structure is reversed with respect to figure
Fig.~\ref{fig:levels}).  Above a certain value of $\theta$, the
component of the magnetic field along $z$, $H_z$, is too weak, such
that the level $M_z=10$ is actually lower in energy than $M_z=-9$, and
$\mathcal{E}_\mathrm{gap}$ corresponds to the difference between the
ortho- and paramagnetic states of the crystal. The kink in
$\mathcal{E}_\mathrm{gap}$ is therefore due to the shift from a
structure of the type of Fig.~\ref{fig:levels}(a) to that of
Fig.~\ref{fig:levels}(b).

In the first case, where the energy gap is between two eigenstates on
the same side of the well, thermal excitation will lead to a small
correction of the activation and Zeeman energies, as the initial state
of the system will have a higher energy than calculated here.  In the
latter case, the thermal energy will lead to an initial projection on
the levels in both wells, leading to a breakdown of the model.


\section{Approximate formulas for the activation and Zeeman energies}
\label{sec:approx}

While an implementation of the rescription of Sec.~\ref{sec:model}
relies on the numerical solution of an eigenvalue system, this can be
done in real time when coupled to a simulation of deflagration or
detonation.  However, it is also useful to have analytical formulas,
which can give insight into the physics governing the processes.  We
therefore derive approximate equations for $\Ea$ and $Q$, for the case
where the external magnetic field is nearly aligned with the easy axis
of the crystal, \ie, $\HT \ll H_z$.  For this purpose, we will also
consider the simplified Hamiltonian
\begin{equation}
  \Hamil = -D \Sp_z^2 - B \Sp_z^4 - g \muB \left( H_z
    \Sp_z + \HT  \Sp_x \right)
  \label{eq:hamil_simple}
\end{equation}
where we have set $\phi = 0$ and neglected the transverse anisotropy
terms.

\subsection{Perturbative approach}

With the exception of the term in $\Sp_x$, the Hamiltonian in
Eq.~(\ref{eq:hamil_simple}) is considered by many authors as the
``unperturbed'' Hamiltonian, the other terms being responsible for a
slight shift of the energy levels and for magnetic tunneling.  This is
also the case for $\HT \ll H_z$, so let us define the unperturbed
Hamiltonian as
\begin{equation}
\Hamil_0 = -D \Sp_z^2 - B \Sp_z^4 - g \muB H_z \Sp_z \, ,
\end{equation}
with the perturbation
\begin{equation}
\Hamil' = - g \muB \HT \Sp_x \, .
\end{equation}
The eigenstates of the unperturbed Hamiltonian will be written as
\begin{equation}
  \Hamil_0 \ket{M_z} = \mathcal{E}_{M}^{(0)} \ket{M_z} ,
\end{equation}
with the unperturbed energy
\begin{equation}
  \mathcal{E}_{M}^{(0)} = -D M_z^2 - B M_z^4 - g \muB H_z M_z.
  \label{eq:E0}
\end{equation}

There is no first order correction to the energy, \ie,
\begin{equation}
  \mathcal{E}_{M}^{(1)} = \braket{M_z | \Hamil' | M_z} = 0,
\end{equation}
since the diagonal elements of $\Sp_x$ are 0.  We thus need to
consider second-order corrections,
\begin{widetext}
\begin{align}
  \mathcal{E}_M^{(2)} &= \sum_{M_z' \neq M_z} \frac{\left|
      \braket{M_z' | \Hamil' | M_z} \right|^2}{\mathcal{E}_{M'}^{(0)}
    - \mathcal{E}_M^{(0)}} = \left[ \frac{\left| \braket{M_z+1 |
          \Hamil' | M_z} \right|^2}{\mathcal{E}_{M+1}^{(0)} -
      \mathcal{E}_M^{(0)}} + \frac{\left| \braket{M_z-1 | \Hamil' |
          M_z} \right|^2}{\mathcal{E}_{M-1}^{(0)} -
      \mathcal{E}_M^{(0)}} \right]
  \nonumber \\
  &= \left( \frac{g \muB \HT}{2} \right)^2 \left\{ \frac{S (S+1)- M_z
      (M_z+1)}{\left( 2M_z+1 \right) \left[ -D -B \left( 2M_z^2 + 2M_z
          +1 \right) \right] - g \muB H_z} \right. \nonumber \\
  &\quad \left. + \frac{S (S+1)- M_z (M_z-1)}{\left( 2M_z-1 \right)
      \left[ D + B \left( 2M_z^2 - 2M_z +1 \right) \right] + g \muB
      H_z} \right\},
\end{align}
such that the total energy is given to second order by
\begin{align}
  \tilde{\mathcal{E}}_M &\equiv  \mathcal{E}_M^{(0)} + \mathcal{E}_M^{(1)}
  + \mathcal{E}_M^{(2)} \nonumber \\
  &= -D M_z^2 -B M_z^4 - g \muB H_z M_z + \left( \frac{g
      \muB \HT}{2} \right)^2  \left\{ \frac{S (S+1)- M_z
      (M_z+1)}{ \left( 2M_z+1 \right) \left[ -D -B \left( 2M_z^2 + 2M_z +1
        \right) \right] -   g \muB
        H_z} \right. \nonumber \\
  & \quad \left.
    + \frac{S (S+1)- M_z
      (M_z-1)}{ \left( 2M_z-1 \right) \left[ D + B \left( 2M_z^2 - 2M_z +1
        \right) \right] +   g \muB
        H_z} \right\}.
    \label{eq:Eperturb}
\end{align}
Explicitly, we have the energy of the initial state $\ket{-S}$,
\begin{equation}
  \tilde{\mathcal{E}}_{-S} = -D S^2 - B S^4 + g \muB H_z S
  +  \frac{S \left( g
      \muB \HT \right)^2}{2 \left( 2S-1
    \right)  \left[ D +B \left( S^2 -2S +1 \right) \right] - 2 g
      \muB H_z},
    \label{eq:Epinit}
\end{equation}
and of the ground state $\ket{S}$,
\begin{equation}
  \tilde{\mathcal{E}}_{S} = -D S^2 - B S^4 - g \muB H_z S
  +  \frac{S \left( g
    \muB \HT \right)^2}{2 \left( 2S-1
    \right)  \left[ D +B \left( S^2 -2S +1 \right) \right] + 2 g
      \muB H_z},
\end{equation}
after inversion of the field.  The Zeeman energy is thus found to be
\begin{equation}
  \tilde{Q} = \tilde{\mathcal{E}}_{-S} -  \tilde{\mathcal{E}}_{S}
  = g \muB H_z S \left\{ 2 + \frac{\left( g
        \muB \HT \right)^2}{\left( 2S-1 \right)^2 \left[ D
        + B \left( S^2 -2S +1 \right) \right]^2 - \left( g
        \muB H_z \right)^2} \right\}.
\end{equation}
\end{widetext}

Calculating the activation energy is more tricky, as it requires
knowledge of the value of $M_z$ for which the energy is maximum.  We
remedy this by considering $M_z$ to be real, and not limited to
integer values.  Using the unperturbed energy, Eq.~(\ref{eq:E0}), we
find
\begin{equation}
  M_{\mathrm{max}} \equiv \max_{M_z} \mathcal{E}_{M}^{(0)} =
  \frac{D}{\left( 3 \gamma \right)^{1/3}} - \frac{\left( \gamma/9
    \right)^{1/3}}{2 B},
\end{equation}
where we have defined
\begin{equation}
\gamma \equiv 9 B^2 g \muB H_z + \sqrt{3 \left[ 8 B^3 D^3 + 27
    B^4 \left( g \muB H_z \right)^2 \right]}.
\end{equation}
We also get that
\begin{equation}
\lim_{B\rightarrow0} M_{\mathrm{max}} = - \frac{g \muB H_z}{2 D}.
\end{equation}
We finally can get the approximate activation energy by substituting
$M_{\mathrm{max}}$ into Eq.~(\ref{eq:Eperturb}) and subtracting the
energy of the initial state [Eq.~(\ref{eq:Epinit})], \ie,
\begin{equation}
  \tilde{\mathcal{E}}_{a} = \tilde{\mathcal{E}}_{M_{\mathrm{max}}} -
  \tilde{\mathcal{E}}_{-S}.
\end{equation}

\begin{figure}
  \includegraphics{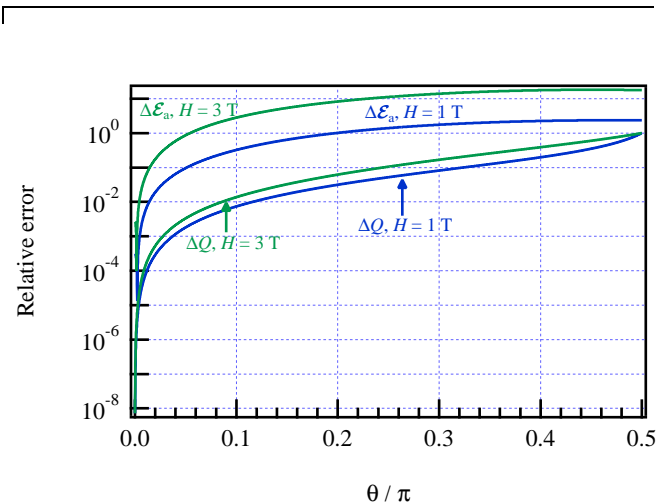}
  \caption{\label{fig:error_perturb}(Color online) Relative error on
    the activation $(\Delta \mathcal{E}_a)$ and Zeeman ($\Delta Q$)
    energies for Mn$_{12}$-acetate calculated using perturbation
    theory for the Hamiltonian Eq.~(\ref{eq:hamil_simple}), for two
    different magnitudes of the field.}
\end{figure}

In Fig.~\ref{fig:error_perturb}, we present
the relative error on the calculation of $\tilde{\mathcal{E}}_{a}$ and
$\tilde{Q}$, as compared to the exact quantum-mechanical calculation,
as presented in Sec.~\ref{sec:model}, but for the
Hamiltonian~(\ref{eq:hamil_simple}).  As expected, the results are in
good agreement for small angles, but a strong deviation is observed as
$\HT$ becomes non-negligible compared to $H_z$.  A much better
approximation is obtained for the Zeeman energy, in great part because
levels close to the top of the barrier are more affected than those at
the bottom of the wells and because of the additional approximation
that is need to determine the value of $M_z$ for which the energy is
maximum.

\subsection{Classical approach}

Following the approach of Maci\`{a} \emph{et al.},\cite{mag:macia09} we
shall now treat the spin of the nanomagnet as a classical vector
$\mathbf{S}$.  By deriving the dependence of the energies with respect
to the orientation of the spin vector, it will be possible determine
the activation and Zeeman energies, following the same method as
prescribed above for the quantum Hamiltonian.

From the Hamiltonian~(\ref{eq:hamil_simple}), we get the classical
formulation of the energy
\begin{align}
  \mathcal{E}^{\mathrm{class}} &= -D S_z^2 - B S_z^4 - g
  \muB \left( H_z S_z + \HT S_x
  \right) \nonumber \\
  &= -D \left(S \cos\alpha \right)^2 - B \left(S \cos\alpha \right)^4
  \nonumber \\
  &\quad - g \muB \left( H_z S \cos\alpha + \HT S \sin\alpha
  \right),
\label{eq:Eclass}
\end{align}
where
$\alpha$ is the angle between the
spin vector and the $z$ axis.  The minimum energy, from which we can
get the orientation of the initial spin vector and that of its ground
state, is therefore found by solving
\begin{align}
  \frac{d\mathcal{E}^{\mathrm{class}}}{d\alpha} &= 2D S^2
  \cos\alpha\sin\alpha + 4 B S^4
  \cos^3\alpha\sin\alpha \nonumber \\
  &\quad - g \muB S \left( - H_z \sin\alpha
    + \HT \cos\alpha \right) = 0.
\label{eq:minim}
\end{align}
Making the assumption that the transverse field $\HT$ is small
compared to the internal anisotropy, see Eq.~(\ref{eq:anisotropy}), we
get that the spin vector will be nearly aligned with the easy axis
($z$), and the external field will introduce only a slight deviation.  This is
indeed what is observed for the full-quantum calculation in
Fig.~\ref{fig:angle}.  The angle $\alpha$ is thus small, such that we
can approximate Eq.~(\ref{eq:minim}) by
\begin{equation}
  \frac{d\mathcal{E}^{\mathrm{class}}}{d\alpha} \approx 2D S^2 \alpha
  + 4 B S^4 \alpha - g
  \muB S \left( - H_z \alpha + \HT \right) = 0,
\end{equation}
and we get
\begin{equation}
\alpha_\mathrm{min} \approx \frac{g \muB \HT}{2DS + 4BS^3 + g
  \muB H_z}.
\label{eq:alpha}
\end{equation}
Thus, the energy of the ground state,
$\mathcal{E}^{\mathrm{class}}_g$, is obtained from
Eq.~(\ref{eq:Eclass}) using Eq.~(\ref{eq:alpha}) for $\alpha$.

The energy of the initial state, $\mathcal{E}^{\mathrm{class}}_i$ is
also obtained from Eq.~(\ref{eq:Eclass}), but using the angle of the
spin vector $\alpha_{\mathrm{min},-}$ in the inverted field,
$\mathbf{H}_-$.  Following the above procedure, we easily find that
\begin{equation}
\alpha_{\mathrm{min},-} = \pi + \alpha_{\mathrm{min}}.
\end{equation}
[The symmetry of
Eq.~(\ref{eq:Eclass}) with respect to the inversion of the external
field can also be used to demonstrate the relation between
$\alpha_{\mathrm{min},-}$ and $\alpha_{\mathrm{min}}$.]  We finally get
\begin{align}
  Q^{\mathrm{class}} &= \mathcal{E}^{\mathrm{class}}_i -
  \mathcal{E}^{\mathrm{class}}_g \nonumber \\
  &= \mathcal{E}^{\mathrm{class}} (\alpha = \pi +
  \alpha_{\mathrm{min}}) - \mathcal{E}^{\mathrm{class}} (\alpha =
  \alpha_{\mathrm{min}})
  \nonumber \\
  &= 2 g \muB S \left( H_z \cos \alpha_{\mathrm{min}} + \HT \sin
    \alpha_{\mathrm{min}} \right).
\end{align}

To calculate the activation energy, we again need to determine the
highest energy the spin vector will have to overcome as its angle goes
from $\alpha_{\mathrm{min},-}$ to $\alpha_{\mathrm{min}}$.  Plotting
Eq.~(\ref{eq:Eclass}) as a function of $\alpha$, one can easily see
that the maximum is around $\alpha \approx 3\pi/2$.  Making the
substitution $\alpha_{\mathrm{max}} = 3\pi/2 + \epsilon$, with
$\epsilon$ a small angle, into Eq.~(\ref{eq:minim}), we have
\begin{widetext}
\begin{equation}
  -2D S^2 \sin\epsilon\cos\epsilon - 4 B S^4
  \sin^3\epsilon\cos\epsilon - g \muB S \left(  H_z \cos\epsilon
    + \HT \sin\epsilon \right)
  \approx -2D S^2 \epsilon - 4 B S^4
  \epsilon^3 - g \muB S \left(  H_z
    + \HT \epsilon \right) = 0.
\end{equation}
Solving for $\epsilon$, we get
\begin{align}
  \alpha_{\mathrm{max}} &= \frac{3\pi}{2} + \frac{1}{2 B^{1/3} S} \left(
    \left\{ -g \muB H_z + \left[ \frac{\left(2DS + g
            \muB \HT \right)^3}{27 B S^3} + \left( g
          \muB H_z \right)^2 \right]^{1/2}
    \right\}^{1/3} \right. \nonumber \\
  &\quad \left. - \left\{ g \muB H_z + \left[
        \frac{\left(2DS + g \muB \HT \right)^3}{27 B S^3}
        + \left( g \muB H_z \right)^2 \right]^{1/2}
    \right\}^{1/3}\right).
\end{align}
From this expression for $\alpha_{\mathrm{max}}$, we find the corresponding
energy $\mathcal{E}^{\mathrm{class}}_{\mathrm{max}}$, leading to
\begin{align}
  \Ea^{\mathrm{class}} &= \mathcal{E}^{\mathrm{class}}_{\mathrm{max}}
  - \mathcal{E}^{\mathrm{class}}_i = \mathcal{E}^{\mathrm{class}}
  (\alpha = \alpha_{\mathrm{max}}) - \mathcal{E}^{\mathrm{class}}
  (\alpha = \pi + \alpha_{\mathrm{min}}) \nonumber \\
  &= -D S^2 \left( \cos^2 \alpha_{\mathrm{max}} - \cos^2
    \alpha_{\mathrm{min}} \right) - B S^4 \left( \cos^4
    \alpha_{\mathrm{max}} - \cos^4 \alpha_{\mathrm{min}} \right)
  \nonumber \\
  &\quad - g \muB S \left[ H_z \left( \cos \alpha_{\mathrm{max}} +
      \cos \alpha_{\mathrm{min}} \right) + \HT \left( \sin
      \alpha_{\mathrm{max}} + \sin \alpha_{\mathrm{min}} \right)
  \right].
\end{align}
\end{widetext}

\begin{figure}
  \centerline{\includegraphics{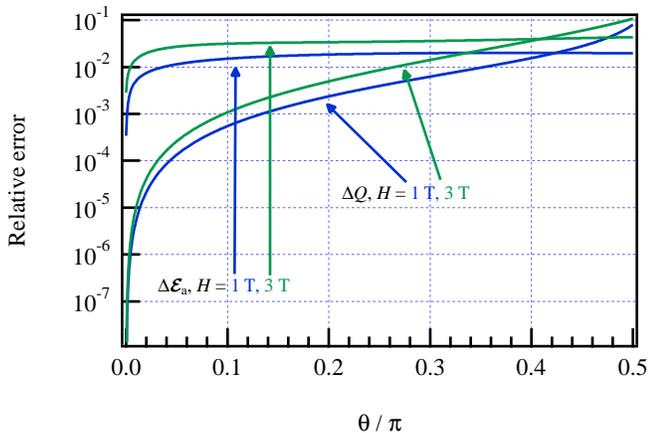}}
  \caption{\label{fig:error_class}(Color online) Relative error on
    the activation $(\Delta \mathcal{E}_a)$ and Zeeman ($\Delta Q$)
    energies for Mn$_{12}$-acetate calculated in the classical
    approximation, based on the Hamiltonian
    Eq.~(\ref{eq:hamil_simple}), for two
    different magnitudes of the field.}
\end{figure}

We once more calculate the relative error with respect to the exact
quantum-mechanical values using the
Hamiltonian~(\ref{eq:hamil_simple}), see Fig.~\ref{fig:error_class}.
The result is markedly better than that obtained using perturbation
theory (Fig.~\ref{fig:error_perturb}), even for greater values of the
angle $\theta$.  This can be easily explained by the fact that the
classical model is based on a much different approximation, namely
that the spin only slightly deviates from being aligned with the easy
($z$) axis.  This gives a validity over a much greater range than what was given by treating
the transverse field $\HT$ as a perturbation.

\section{Anisotropic properties of magnetic deflagration and
  detonation}
\label{sec:properties}

\subsection{Magnetic deflagration}

\begin{figure}
\centerline{\includegraphics[width=2.0in]{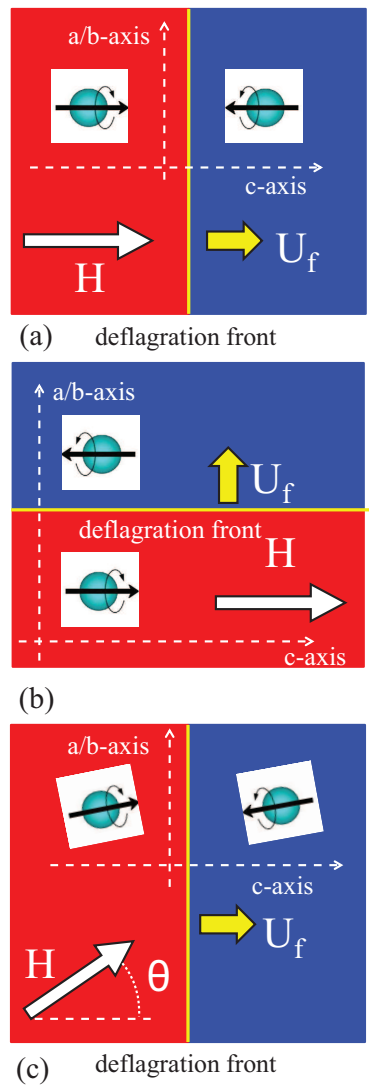}}
\caption{\label{fig:schematic}(Color online) Schematic of the
  deflagration front geometry in the crystal of nanomagnets for the
  following three cases: (a) the external magnetic field and the front
  propagation are parallel to the easy axis; (b) the magnetic field is
  parallel, but the front propagation is perpendicular to the easy
  axis; (c) front propagation is parallel to the easy axis, but the
  field is at some angle to the axis.}
\end{figure}

In this subsection, we investigate the magnetic deflagration speed for
an arbitrary angle between the magnetic field and the easy axis; the
next subsection will be devoted to magnetic detonation. We stress here
that the propagation of magnetic deflagration involves four important
vector values: the magnetic field intensity $\mathbf{H}$, the
magnetization $\mathbf{M}$, the temperature gradient $\nabla T$, and
the heat flux $\hat{\kappa}\nabla T$ (with $\hat{\kappa}$ being the
tensor of thermal conduction). The latter two are in general not
parallel because of the crystal anisotropy. Among these values, the
temperature gradient $\nabla T$ determines the direction of front
propagation, while the magnetic field intensity $\mathbf{H}$ and
magnetization $\mathbf{M}$ specify the activation energy and the
Zeeman energy of the spin reversal as discussed in the previous
sections. We stress that the vectors $\mathbf{H}$ and $\mathbf{M}$ are
not related to the direction of the deflagration front velocity, but
influence the absolute value of that velocity. We also point out that
1) the direction of the magnetic field $\mathbf{H}$ is controlled by
the experimental set-up; 2) the direction of the magnetization
$\mathbf{M}$ correlates strongly with the easy axis (c-axis) of the
crystal (see the calculations above and Fig.~\ref{fig:angle}); 3) the
direction of the temperature gradient $\nabla T$ and front propagation
are determined by the ignition conditions, \eg, by surface acoustic
waves;\cite{mag:hernandez-minguez05} and 4) the direction of the heat
flux $\hat{\kappa}\nabla T$ results from the anisotropic thermal
conduction of the crystal. The different directions defined by these
four vectors open a wide parameter space for experimental studies of
anisotropic crystal properties, both magnetic and thermal.  As an
example, Fig.~\ref{fig:schematic}
illustrates some possibilities of the magnetic deflagration geometry
[because of the small factor $E=10^{-2}\ \mathrm{K}$ in the
Hamiltonian for Mn$_{12}$-acetate (see Tab.~\ref{tab:params}) the
difference between the a and b crystal axes is
minor]. Figure~\ref{fig:schematic}(a) shows the commonly investigated
case of a deflagration front propagating along the easy axis with the
magnetic field and magnetization aligned along the same axis. In
Fig.~\ref{fig:schematic}(b), the magnetic field points along the easy
axis, but the magnetic deflagration front propagates along the hard
axis (axis a or b). Obviously, both the activation and Zeeman energies
are the same for the geometries of Fig.~\ref{fig:schematic}(a) and
(b); but the deflagration speed is different because of different
thermal conduction along the easy and hard axes. In particular, by
comparing the magnetic deflagration speed for these two geometries,
${U_{f(a,b)}}$ and ${U_{f(c)}}$, one can measure the ratio of the
thermal conduction coefficients ${\kappa_{a,b}}/{\kappa_c}$
quantitatively as
${\kappa_{a,b}}/{\kappa_c}=[{U_{f(a,b)}}/{U_{f(c)}}]^{2} $.  Finally,
Fig.~\ref{fig:schematic}(c) shows the geometry with the front
propagating along the easy axis, but with the magnetic field directed
at some arbitrary angle to the axis. In this section we focus on the
geometry of Fig.~\ref{fig:schematic}(c). For large magnetization
values, $M \sim H$, this geometry involves refraction of the magnetic
field at the deflagration front. Still, for the crystals of
nanomagnets used in the experimental studies so far, the magnetization
is small, $M \ll H$, and the refraction effects may be neglected.  In
principle, one may consider an even more general geometry than that
shown in Fig.~\ref{fig:schematic}(c) with both the magnetic field and
the front speed aligned at some angle to the easy axis. However, at
present there is no quantitative experimental data for the ratio
${\kappa_{a,b}}/{\kappa_c}$; therefore, such a general case involves
unidentified parameters and, without proper experimental support, it
may be considered only as an hypothetical study. A qualitative
comparison of the coefficients of thermal conduction along different
axes ${\kappa_{a,b,c}}$ was performed in Ref.~\onlinecite{mag:velez12}
for crystals of $\textrm{Gd}_{5}\textrm{Ge}_{4}$, leading to the
evaluation that ${\kappa_{a}}> {\kappa_{b}}>{\kappa_c}$. Assuming the
same tendency for $\textrm{Mn}_{12}$-acetate, one should expect that
the thermal anisotropy somewhat moderates the strong effects of
magnetic anisotropy obtained below. Still, a noticeable influence of
thermal anisotropy is unlikely since the difference between the
coefficients of thermal conduction ${\kappa_{a,b,c}}$ is presumably
only by a numerical factor of order of unity and the magnetic
deflagration speed depends rather weakly on ${\kappa}$ as
$U_{f}\propto \sqrt{{\kappa}}$. In contrast to that, we show below
that magnetic anisotropy leads to variations of the magnetic
deflagration speed by two orders of magnitude.

\begin{figure}
\centering
\includegraphics[width=0.48\textwidth]{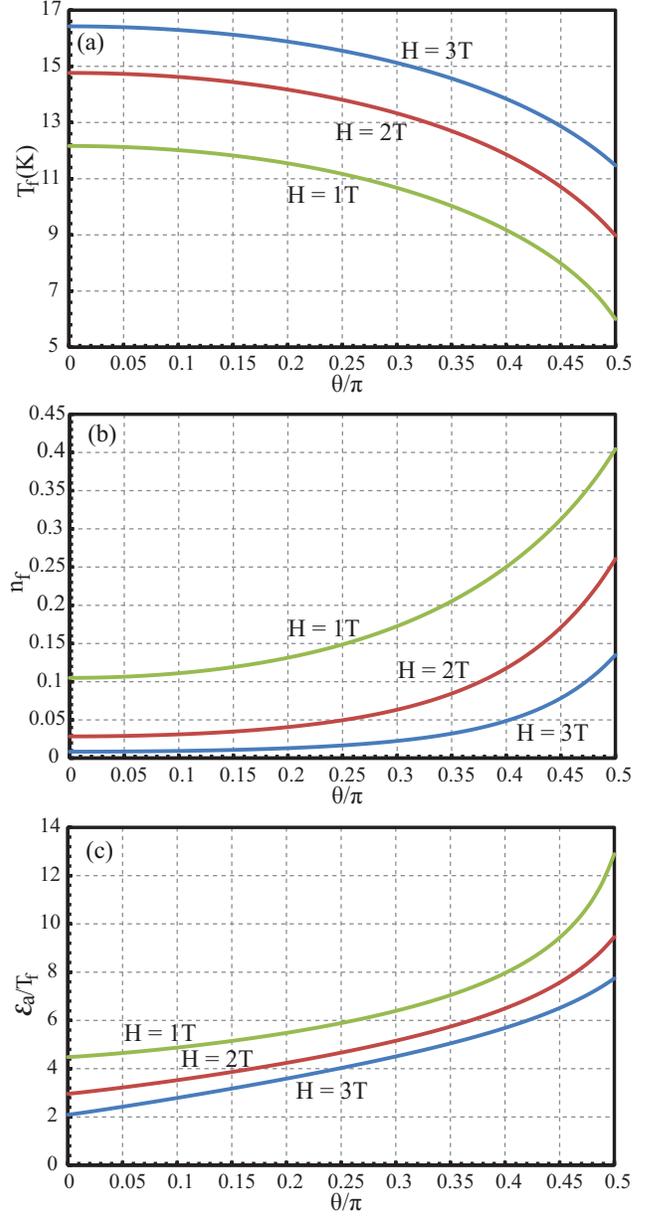}
\caption{\label{fig:Par_vs_Angle}(Color online) The parameters of the
  magnetic deflagration vs the angle between the magnetic field and
  the easy axis: (a) final temperature, (b) final concentration of the
  metastable molecules, (c) scaled activation energy.}
\end{figure}

\begin{figure}
\centering
\includegraphics[width=0.48\textwidth]{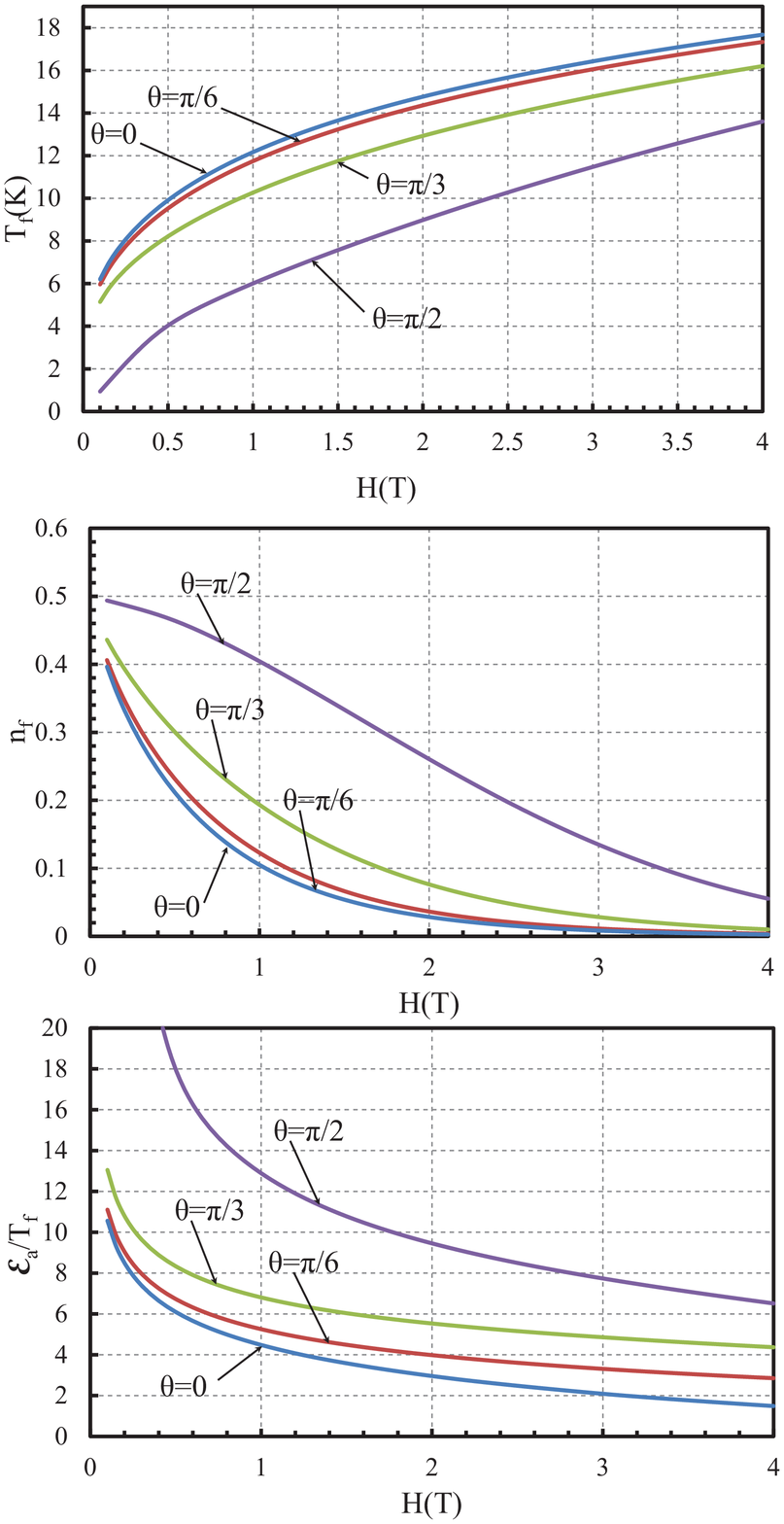}
\caption{\label{fig:Par_vs_field}(Color online) The parameters of the
  magnetic deflagration vs the magnetic field: (a) final temperature,
  (b) final concentration of the metastable molecules, (c) scaled
  activation energy.}
\end{figure}

Within the geometry of Fig.~\ref{fig:schematic}(c), the governing
equations for magnetic deflagration
are\cite{mag:garanin07,mag:modestov11a}
\begin{equation}
  \label{Energy release}
  \frac{\partial \mathcal{E}}{\partial t} = \nabla\cdot\left( \kappa
    \nabla \mathcal{E} \right) - Q\frac{\partial n}{\partial t}
\end{equation}
and
\begin{equation}
  \label{concentration governing}
  \frac{\partial n}{\partial t} = -\frac{1}{\tau_{R}}
  \exp\left(-\frac{\mathcal{E}_{a}}{T}\right)
  \left[n-\frac{1}{\exp\left(Q/T\right)+1}\right],
\end{equation}
where $\mathcal{E}$ is the phonon energy, $T$ is temperature, $n$ is
the fraction of molecules in the metastable state (\ie, normalized
concentration), $\tau_{R}$ is the coefficient of time dimension
characterizing the kinetics of the spin switching.  We also take into
account here the possibility of a non-zero final fraction of molecules
in the metastable state in the case of relatively low heating (low
Zeeman energy), which has been termed ``incomplete magnetic burning''
in Ref.~\onlinecite{mag:garanin07}. This fraction is given
by\cite{mag:garanin07,mag:modestov11a}
\begin{equation}
  \label{metastable concentration}
  n_{f}=\frac{1}{\exp\left(Q/T\right)+1} ,
\end{equation}
which is (obviously) taken into account in Eq.~(\ref{concentration governing}); here the
label $f$ refers to the final state of the system after the
avalanche. As we can see from Figs.~\ref{fig:Par_vs_Angle} and
\ref{fig:Par_vs_field}, the concentration $n_{f}$ cannot be neglected
in the case of a small magnetic field and/or strong misalignment with
the c-axis.  The phonon energy and crystal temperature in
Eqs.~(\ref{Energy release}) and (\ref{concentration governing}) are
related according to\cite{mag:garanin07,kittel63}
\begin{equation}
\label{energy}
\mathcal{E} = \frac{A k_{B}}{\alpha+1}\left(
  \frac{T}{\Theta_{D}}\right)^{\alpha}T,
\end{equation}
where $A=12\pi^4/5$ corresponds to the simple crystal model, $k_{B}$
is the Boltzmann constant, $\alpha$ is the problem dimension (we take
$\alpha=3$, as we consider the 3D case), $\Theta_{D}$ is the Debye
temperature, with $\Theta_{D} = 38$~K for $\textrm{Mn}_{12}$
acetate. The thermal conduction may also depend on temperature;
Refs.~\onlinecite{mag:garanin07,mag:modestov11a} considered the
dependence in the form $\kappa\propto T^\beta$ with the parameter
$\beta$ within the range 0 to 13/3. Below we show that the case of
constant thermal conduction, \ie, $\beta=0$, gives the best fit to the
experimental data.\cite{mag:suzuki05}

We consider the stationary solution to Eqs.~(\ref{Energy release}) and
(\ref{concentration governing}) for a planar magnetic deflagration
front propagating with constant velocity $U_{f}$ along the $z$-axis
(the easy axis). In the reference frame of the front,
Eqs. (\ref{Energy release}) and (\ref{concentration governing}) reduce
to
\begin{equation}
  \label{Energy release2}
  U_{f}\frac{d}{dz}\left( \mathcal{E}+Q n\right) = \frac{d}{dz}\left(
    \kappa\frac{d \mathcal{E}}{dz}\right),
\end{equation}
\begin{equation}
  \label{concentration governing2}
  U_{f}\frac{dn}{dz} = -\frac{1}{\tau_{R}}
  \exp{\left(-\frac{\mathcal{E}_{a}}{T}\right)}
  \left[n-\frac{1}{\exp\left(Q/T\right)+1}\right].
\end{equation}
The boundary conditions for the system are determined by the initial
energy $\mathcal{E}_{0}$ (temperature $T_{0}$) far ahead of the front,
and the final energy $\mathcal{E}_{f}$ (temperature $T_{f}$) far behind
the front. The initial and final energies (temperatures) are related
by the condition of energy conservation $\mathcal{E}_{0}+Q n_{0} =
\mathcal{E}_{f}+Q n_{f}$, or
\begin{eqnarray}
\label{final temp}
\frac{AT_{0}^{\alpha+1}}{\left(\alpha+1\right)\Theta_{D}^{\alpha}}+Q\left(
  1-\frac{1}{\exp\left(Q/T_{0}\right)+1}\right) \nonumber \\
=\frac{AT_{f}^{\alpha+1}}{\left(\alpha+1\right)\Theta_{D}^{\alpha}}+\frac{Q}{\exp\left(Q/T_{f}\right)+1},
\end{eqnarray}
which follows from Eq.~(\ref{Energy release2}). In particular,
our calculations use a low initial temperature, $T_{0}=0.2$~K,
which allows reducing Eq.~(\ref{final temp}) to the simpler form
\begin{equation}
\label{final temp2}
 \frac{AT_{f}^{\alpha}}{\left(\alpha+1\right)\Theta_{D}^{\alpha}}=
 \frac{Q/T_{f}}{1+\exp\left(-Q/T_{f}\right)}.
\end{equation}
We calculate final temperature $T_{f}$ and the final molecule fraction
in the metastable state $n_{f}$ numerically for different strengths
and inclinations of the magnetic field; the results obtained are
presented in Figs.~\ref{fig:Par_vs_Angle} and \ref{fig:Par_vs_field}
together with the scaled activation energy $\mathcal{E}_{a}/{T}_{f}$,
which plays an important role for the deflagration front dynamics. As
we can see, the temperature $T_{f}$ increases with the field and
decreases with the angle; the scaled activation energy
$\mathcal{E}_{a}/{T}_{f}$ decreases with the field and increases with
the angle. Still, this decrease/increase is not dramatic; for example,
for $H = 1\ \textrm{T}$, the temperature changes from $12.2\
\textrm{K}$ to $6.0\ \textrm{K}$ and the scaled activation energy from
$4.5$ to $11.8$ as the angle $\theta$ varies from $0$ to $\pi / 2$. We
will see below that the variations of the deflagration speed are much
stronger because the speed is sensitive to both the final temperature
and the scaled activation energy.


A qualitative understanding of the magnetic deflagration speed may be
obtained from the Zeldovich-Frank-Kamenetsky
theory, from which we have the expression\cite{comb:zeldovich85,mag:garanin07}
\begin{equation}
  \label{asymptotic velocity}
  U_{f} = \sqrt{\frac{\kappa_{f}}{Z\tau_{R}}}
  \exp{\left(-\frac{\mathcal{E}_{a}}{2T_{f}}\right)},
\end{equation}
where $Z$ is the Zeldovich number,
\begin{equation}
\label{Zeldovich number}
Z=\frac{\mathcal{E}_{a}}{T_{f}}\frac{Q\left( 1 - n_{f}\right)}{C_{f}T_{f}}
\sim \frac{1}{\left(\alpha+1\right)}\frac{\mathcal{E}_{a}}{T_{f}},
\end{equation}
and $C_{f}\equiv(d \mathcal{E}/dT)_{f}$ is the heat capacity in the
heated crystal. The final relation in Eq.~(\ref{Zeldovich number})
becomes an accurate equality for the case of complete magnetic
burning, $n_{f} \ll 1$.  The Zeldovich-Frank-Kamenetsky theory, giving
the speed [Eq.~(\ref{asymptotic velocity})], holds only for large
values of the Zeldovich number $Z\gg1$. Such large values are common
in combustion problems,\cite{comb:law06,comb:bychkov00} but rather
unusual for magnetic deflagration. As we can see from
Figs.~\ref{fig:Par_vs_Angle} and \ref{fig:Par_vs_field}, the
Zeldovich-Frank-Kamenetsky theory may be applied to magnetic
deflagration only for the cases of sufficiently low field and high
angles between the magnetic field and the easy axis approaching $\pi
/2$.  In the case of a moderate Zeldovich number, as often encountered
in magnetic deflagration, the deflagration speed may be calculated
numerically on the basis of Eqs. (\ref{Energy release2}) and
(\ref{concentration governing2}) using the numerical method of
Refs.~\onlinecite{mag:modestov11a,comb:modestov09}.

\begin{figure}
\centering
\includegraphics[width=3.3in]{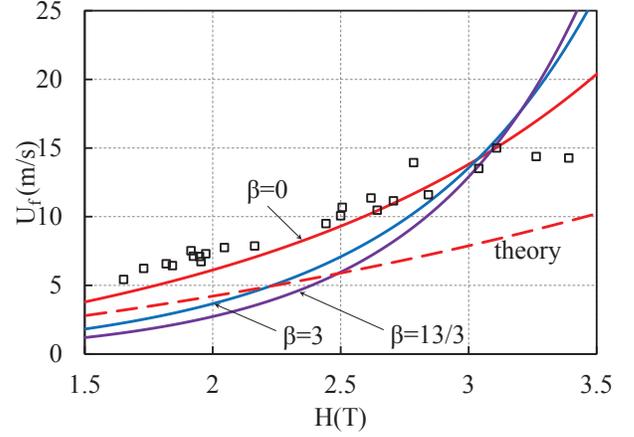}
\caption{\label{fig:vel_compare}(Color online) Comparison of the
  experiments and numerical calculations for the magnetic deflagration
  speed versus the applied magnetic field. The markers show the
  experimental data of Ref.~\onlinecite{mag:suzuki05}. The solid
  lines present the numerical solutions for different temperature
  dependencies of the thermal conduction coefficient with $\beta=0;
  \,3; \,13/3$ and $\kappa_{f}/\tau_{R}=207\,
  \textrm{m}^2/\textrm{s}^2$ providing the best fit of the
  experimental data. The dashed line presents the analytical theory,
  Eq.~(\ref{asymptotic velocity}), plotted for $\beta=0$ and
  $\kappa_{f}/\tau_{R}=207 \,\textrm{m}^2/\textrm{s}^2$.}
\end{figure}

We point out that the problem contains a number of parameters whose
experimental measurement still remain a challenging task, such as the
thermal conduction $\kappa_{f}$ and the coefficient of time dimension
characterizing spin-switching $\tau_{R}$. The temperature dependence
of thermal conduction $\kappa\propto T^\beta$ is also unclear, with
the factor $\beta$ treated as a free parameter in
Refs.~\onlinecite{mag:garanin07,comb:modestov09} changing within the
range of $0<\beta<13/3$.  We suggest here choosing particular values
of the unknown parameters by comparing numerical results to the
experimental data\cite{mag:suzuki05} obtained for the magnetic field
aligned along the easy axis. Figure~\ref{fig:vel_compare} presents the
magnetic deflagration speed versus the magnetic field calculated for
different values of $\kappa_{f}/\tau_{R}$ and $\beta$. Comparison to
the experimental data suggests the parameter values
$\kappa_{f}/\tau_{R}=207\ \textrm{m}^2/\textrm{s}^2$ and $\beta=0$,
which provide the best fit for the experimental results (red line) and
which we use in the following for investigating the anisotropic
properties of magnetic deflagration. The method of least squares was
used to fit the data. As we can see in Fig.~\ref{fig:vel_compare}, a
strong temperature dependence of the thermal conduction $\kappa\propto
T^\beta$ with $\beta=3,\, 13/3$ leads to an excessively strong
dependence of the deflagration speed on the magnetic field, which does
not reproduce the experimental data properly.
Figure~\ref{fig:vel_compare} shows also the analytical predictions of
the Zeldovich-Frank-Kamenetsky theory, Eq.~(\ref{asymptotic
  velocity}), plotted by the dashed line for the same parameters
$\kappa_{f}/\tau_{R}=207 \,\textrm{m}^2/\textrm{s}^2$ and $\beta=0$ as
the numerical solution. As we can see, the analytical theory provides
only qualitative predictions in the experimentally interesting
parameter range.


\begin{figure}
\centering
\includegraphics[width=3.3in]{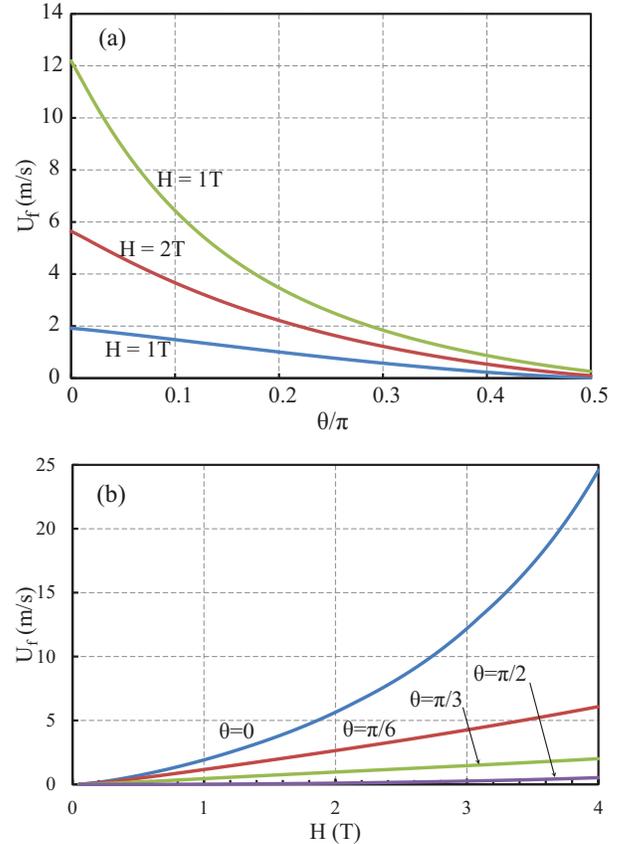}
\caption{ \label{fig:front vel}(Color online) Magnetic deflagration
  speed (a) versus the angle between the magnetic field and the easy
  axis, and (b) versus the magnetic field strength for different
  angles.}
\end{figure}
The numerical results for the magnetic deflagration speed are
presented in Fig.~\ref{fig:front vel}: (a) versus the angle between
the magnetic field and the easy axis for different strength of the
magnetic field; (b) versus the magnetic field strength for different
values of the angle. All plots in Fig.~\ref{fig:front vel} demonstrate
the same tendencies -- monotonic increase of the deflagration speed
with the field and strong decrease with the angle. The tendencies are
qualitatively the same as one had for the final temperature; still,
they are much more dramatic for the deflagration speed. In particular,
for a field strength of $H=3\, \mathrm{T}$ we find the deflagration
speed $U_{f}=12.2\, \mathrm{m/s}$ for for the magnetic field aligned
along the easy axis ($\theta = 0$), a much smaller speed $U_{f}=2.6\,
\mathrm{m/s}$ for $\theta = \pi /4$ and a negligible value
$U_{f}=0.27\ \mathrm{m/s}$ for the magnetic field perpendicular to the
easy axis with $\theta = \pi /2$. Thus we obtain a magnetic
deflagration speed almost two orders of magnitude smaller for the
magnetic field directed along the hard axis in comparison with that
directed along the easy axis. Here we stress that the difference in
the deflagration speed in our study comes only from modifications in
the activation energy and Zeeman energy while the thermal conduction
coefficient remains the same. This is different from the experimental
studies of Ref.~\onlinecite{mag:velez12} for
$\textrm{Gd}_{5}\textrm{Ge}_{4}$ where the deflagration speed changes
both because of misalignment of the magnetic field and thermal
conduction simultaneously. As a result, the geometry suggested here
provides better conditions for investigating quantum-mechanical
properties of the nanomagnets (\ie, magnetic anisotropy) and thermal
properties of the crystals separately.  We also stress that the
present numerical results rely on the available models for the
nanomagnet Hamiltonian for Mn$_{12}$-acetate;\cite{mag:delbarco05} by
modifying the coefficients in the Hamiltonian one comes to other
numerical values for the magnetic deflagration speed. The present work
may also serve for solving the inverse problem: by comparing the
numerical predictions to future refined experiments one may adjust the
coefficients in the Hamiltonian for nanomagnets.

\subsection{Magnetic detonation}

\begin{figure}
\centering
\includegraphics[width=3.3in]{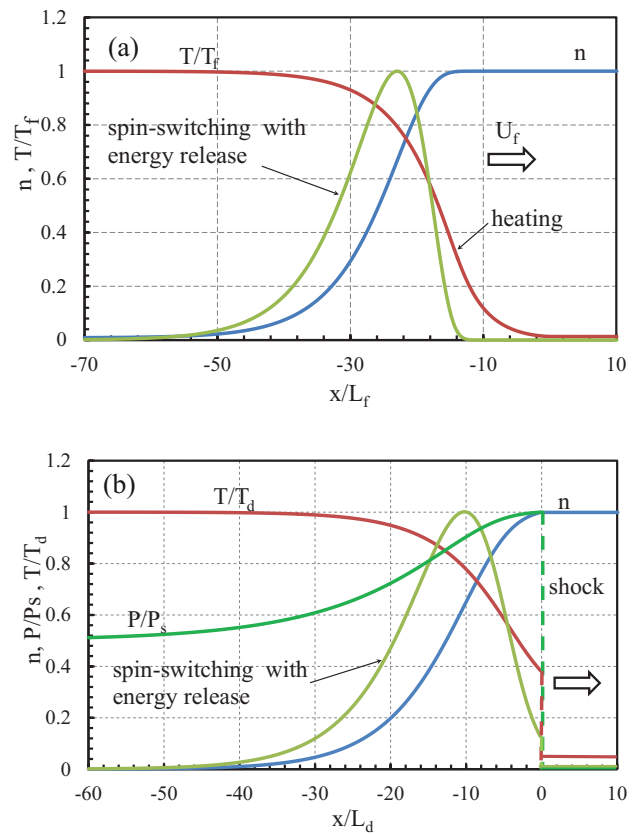}
\caption{\label{fig:Def&det}(Color online) Stationary profiles of the
  scaled temperature $T$, fraction of molecules in the metastable
  state $n$, pressure $P$ (for detonation), and scaled energy release
  for (a) deflagration and (b) detonation for $H=3\,\mathrm{T}$.
  The characteristic length scales are $L_{f}\equiv
  \kappa/U_{f}=1.4\,\mathrm{\mu m}$ for magnetic deflagration and
  $L_{0}\equiv c_{0}\tau_{R}\sim0.2\,\mathrm{mm}$ for magnetic
  detonation.}
\end{figure}
The same method may also be used to investigate anisotropic properties
of magnetic detonation. In contrast to deflagration, magnetic
detonation propagates due to a leading shock wave preheating the
initially cold crystal, see Fig.~\ref{fig:Def&det} for typical
profiles of temperature, pressure and fraction of molecules in the
metastable state. For comparison, in magnetic deflagration, preheating
happens due to thermal conduction, which is negligible for the fast
process of magnetic detonation.  Another important feature of
Fig.~\ref{fig:Def&det} (a) is that the preheating zone for magnetic
deflagration is comparable by width to the zone of spin switching and
energy release at $H = 3\, \textrm{T}$. This is qualitatively
different from the analytical Zeldovich-Frank-Kamenetsky deflagration
model,\cite{mag:garanin07,comb:zeldovich85} which assumes a wide
preheating region and an extremely narrow zone of energy release. We
also point out that magnetic detonation is noticeably different from
the common detonation model (the Zeldovich-von~Neumann-Doring model)
employed in combustion science. In particular, the combustion model
involves a strong delay of the energy release behind the leading
shock.\cite{comb:law06} In contrast to that, in magnetic detonation
the spin switching and energy release start directly at the leading
shock at $H = 3\, \textrm{T}$. The most important properties of
magnetic detonation propagating along the easy axis have been studied
in Ref.~\onlinecite{mag:modestov11b}. In particular,
Ref.~\onlinecite{mag:modestov11b} has demonstrated that magnetic
detonation is ultimately weak in comparison with common combustion
detonations\cite{comb:law06} and, therefore, it propagates with a
velocity only slightly exceeding the sound speed ($c_{0} \approx
2000\, \textrm{m/s}$ for Mn$_{12}$-acetate). As a result, the magnetic
detonation speed does not depend on the direction of the magnetic
field. Unlike that, other properties of magnetic detonation are quite
sensitive to the energy release in the spin switching and hence to the
magnetic field direction. This dependence concerns first of all the
temperature behind the leading shock (label s), which may be
calculated as\cite{mag:modestov11b}
\begin{figure}
\centering
\includegraphics[width=3.3in]{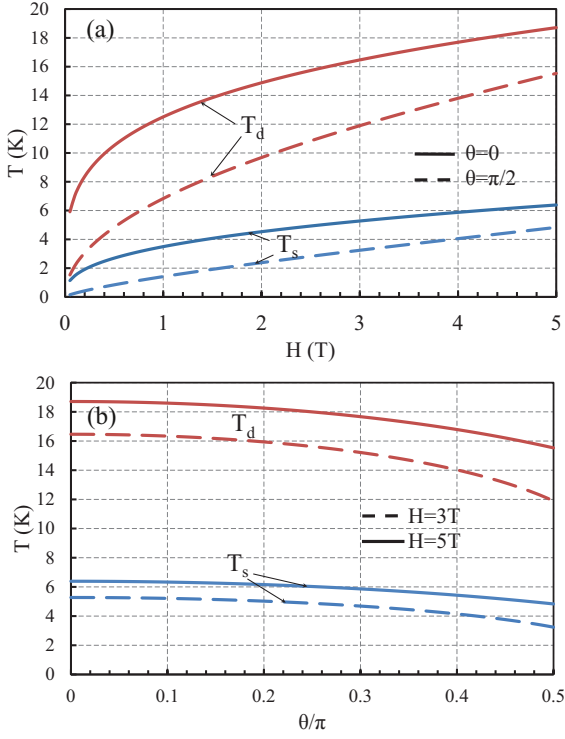}
\caption{\label{fig:T vs. field_angle}(Color online) Temperature
  behind the leading shock and behind the detonation front (a) versus
  the external magnetic field, and (b) versus the angle between the
  magnetic field and the easy axis}
\end{figure}
\begin{equation}
\label{detonation T}
T_{s}^{\alpha + 1} = \left( {\alpha + 1} \right)\left( {m + 1} \right) \frac{2\Theta_{D}^{\alpha
}}{3A k_{B}c_{0}} \left(\frac{2\Gamma Q}{m + 1}\right) ^{3/2},
\end{equation}
where $\Gamma \approx 2$ is the Gruneisen coefficient, and the factor
$m\approx 4$ characterizes the elastic contribution to the pressure $P
\propto (\rho/\rho_{0})^{m}-1$, where
$\rho_{0}\approx1.38\times10^{3}\,\textrm{kg/m}^3$ is the initial
density of the crystal, see Ref.~\onlinecite{mag:modestov11b} for
details. The temperature behind the magnetic detonation front (labeled
d) depends also on the Zeeman energy release as
\begin{equation}
\label{detonation T-d}
T_{d}^{\alpha + 1} = \left( {\alpha + 1} \right) \frac{\Theta_{D}^{\alpha
}}{A k_{B}} \left[ {Q  +  \frac{m + 1}{12c_{0}}  \left(\frac{2\Gamma Q}{m + 1}\right) ^{3/2}}\right].
\end{equation}
The characteristic times of spin switching in magnetic detonation at
the shock, $\tau_{s} \sim \tau_{R}
\exp\left(\mathcal{E}_{a}/T_{s}\right)$, and at the final detonation
temperature, $\tau_{d} \sim \tau_{R}
\exp\left(\mathcal{E}_{a}/T_{d}\right)$, are also strongly influenced
by the direction of the magnetic field.  The anisotropic dependence of
the temperature on the angle between the magnetic field and the easy
axis is presented in Fig.~\ref{fig:T vs. field_angle}.
%
\begin{figure}
\centering
\includegraphics[width=3.3in]{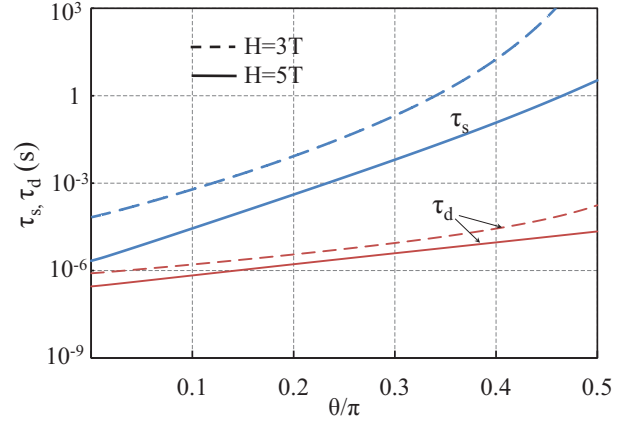}
\caption{\label{fig:Tau_vs_angle}(Color online) The characteristic
  spin-switching time at the shock and at the final detonation
  temperature versus the angle between the magnetic field and the easy
  axis.}
\end{figure}
Similarly to deflagration, the temperature in magnetic detonation
exhibits noticeable, though not dramatic, decrease with the angle
between the magnetic field and the easy axis. For example, for $H =
5\, \textrm{T}$, the temperature just behind the shock changes from
$T_{s}= 6.38 \, \textrm{K}$ at $\theta = 0$ to $T_{s}= 4.9 \,
\textrm{K}$ at $\theta = \pi / 2$; the resulting temperature behind
the magnetic detonation front changes from $T_{d}= 18.7 \, \textrm{K}$
at $\theta = 0$ to $T_{d}= 15.7 \, \textrm{K}$ at $\theta = \pi /
2$. However, these moderate modifications of temperature, together
with respective modifications of the activation energy, lead to
dramatic changes in the characteristic spin-switching time at the
shock, $\tau_{s}$, and at the final detonation temperature,
$\tau_{d}$, as shown in Fig.~\ref{fig:Tau_vs_angle}. For example, for
the same magnetic field strength as used in the above example, $H =
5\, \textrm{T}$, we find the reversal time behind the leading shock
$\tau_{s}= 2.2\times10^{-6} \, \textrm{s}$ at $\theta = 0$ and
$\tau_{s}= 2.3 \, \textrm{s}$ at $\theta = \pi / 2$; thus we observe
variations of the reversal time by six orders of magnitude. Such an
increase of the spin-reversal time makes the magnetic detonation front
unrealistically wide at large angles so that magnetic detonation
becomes impossible for noticeable misalignment between the external
magnetic field and the easy axis.  The characteristic spin-switching
time at the final temperature $T_{d}$ and the external field $H = 5\,
\textrm{T}$ changes from $\tau_{d}= 2.8\times10^{-7} \, \textrm{s}$ at
$\theta = 0$ to $\tau_{d}= 2\times10^{-5} \, \textrm{s}$ at $\theta =
\pi / 2$.  Note that in Fig.~\ref{fig:T vs. field_angle} we consider
larger values of the external magnetic field than those used in the
magnetic deflagration experiments. As pointed out in
Ref.~\onlinecite{mag:modestov11b}, moderate magnetic fields lead to a
quite large thickness of the magnetic detonation front, $\sim
c_{0}\tau_{s}$, much larger than the typical sample size unless the
magnetic detonation is formed at a specific resonant field
characterizing nanomagnets.\cite{mag:decelle09} Investigation of spin
avalanches at the resonant field requires further work beyond the
scope of the present paper.

\section{Summary}
\label{sec:summary}

In this paper, we have investigated anisotropic properties of spin
avalanches in crystals of nanomagnets propagating in the form of
pseudo-combustion fronts -- magnetic deflagration and detonation. In
general, the anisotropy is expected to be of two types: magnetic and
thermal. We have focused here on the magnetic anisotropy related to
the misalignment of the external magnetic field and the easy axis of
the crystal. The thermal anisotropy is not considered since at present
there is no sufficient experimental data for such a study.

The magnetic anisotropy affects primarily two values of the key
importance for the magnetic deflagration and detonation dynamics --
the activation energy and the Zeeman energy. Here, we calculated the
activation and Zeeman energies as a solution to the quantum-mechanical
problem of a single nanomagnet of $\textrm{Mn}_{12}$-acetate placed in
the external magnetic field, which is then reversed. We demonstrated
strong dependence of the activation and Zeeman energies on the
strength and direction of the external magnetic field.

We obtained that, because of this strong dependence, the magnetic
deflagration speed is quite sensitive to the direction of the magnetic
field too. In particular, we found that the magnetic deflagration speed
may decrease by two orders of magnitude for the magnetic field aligned
along the hard crystal axis instead of the easy one.

In contrast to magnetic deflagration, the magnetic detonation speed is
determined mostly by the sound speed in the crystal and, hence, does
not depend on the direction of the magnetic field. At the same time,
other properties of magnetic detonation, such as the temperature
behind the leading shock and for completed spin reversal, and the
characteristic time of spin switching, demonstrate a strong
anisotropy.

\begin{acknowledgments}
  Financial support from the Swedish Research Council and the Faculty
  of Natural Sciences, Ume{\aa} University is gratefully acknowledged. The
  authors thank Myriam Sarachik for useful discussions.
\end{acknowledgments}


%

\end{document}